\documentclass[aps,preprintnumbers,amsmath,amssymb,showpacs,showkeys,twocolumn,floatfix]{revtex4}

\usepackage{graphicx}   
\usepackage{units}
\usepackage{natbib}
\renewcommand{\[}{\begin{equation}}
\renewcommand{\]}{\end{equation}}
\renewcommand{\(}{\begin{equation*}}
\renewcommand{\)}{\end{equation}*}
\begin{document}

\def\ket#1{{%
  \ifmmode |\,#1\,\rangle \else $|\,#1\,\rangle$\fi}}
\def\bra#1{{%
  \ifmmode \langle\,#1\,| \else $\langle\,#1\,|$\fi}}

\title{Comparing contact and dipolar interaction in a \uppercase{B}ose-\uppercase{E}instein condensate}
\author{Axel Griesmaier}
\email{a.griesmaier@physik.uni-stuttgart.de}
\author{J\"urgen Stuhler}
\author{Tobias Koch}
\author{Marco Fattori}
\author{Tilman Pfau}
\author{Stefano Giovanazzi}
\affiliation{5. Physikalisches Institut, Universit\"at Stuttgart, 70550 Stuttgart, Germany.}
\homepage{http://www.pi5.uni-stuttgart.de}
\date{\today}

\begin{abstract}
We have measured the relative strength $\varepsilon_{dd}$ of the
magnetic dipole-dipole interaction compared to the contact
interaction in a chromium Bose-Einstein condensate. We analyze the
asymptotic velocities of expansion of a dipolar chromium BEC with
different orientations of the atomic magnetic dipole moments. By
comparing them with numerical solutions of the hydrodynamic
equations for dipolar condensates, we are able to determine
$\varepsilon_{dd}=0.159\pm0.034$ with high accuracy. Since the
absolute strength of the dipole-dipole interaction is known exactly,
the relative strength of the dipole-dipole interaction can be used
to determine the s-wave scattering length
$a=5.08\pm1.06\cdot10^{-9}$m$= 96\pm20\;a_0$ of $\rm ^{52}Cr$. This
is fully consistent with our previous measurements on the basis of
Feshbach resonances.
\end{abstract}



\maketitle
Gaseous Bose-Einstein condensates (BEC) with dipole-dipole
interaction (DDI) have become a fast growing field of theoretical
and experimental interest. Many new exciting phenomena are expected.
Due to the anisotropic character of the DDI, most of them depend
strongly on the symmetry of the trap. The expected phenomena range
from modifications of the ground state wave
function~\cite{Goral:2000,Yi:2000}, the
expansion~\cite{Giovanazzi:2003a,Yi:2003,Giovanazzi:2006}, the
excitation
spectrum~\cite{Santos:2003,Goral:2002,ODell:2004,Pedri:2005}, and
stability criteria~\cite{Santos:2000,Goral:2000,Goral:2002} to the
occurrence of new quantum phases in optical
lattices~\cite{Goral:2002a} and dramatic influence on the formation
of vortices and vortex lattices~\cite{Cooper:2005,Rezayi:2005}.
Dipolar BECs are now also discussed in the context of spinor
condensates~\cite{Santos:2005a, Diener:2005}, where the combination
of large spin and magnetic moment leads to new effects like the
conversion of spin into angular
momentum~\cite{Kawaguchi:2005,Santos:2005a}. For all of these
phenomena, the relative strength of the DDI compared to the contact
interaction is a
very important parameter.\\
In two recent publications, we have reported on the generation of a
BEC of chromium atoms ($\rm ^{52}Cr$)~\cite{Griesmaier:2005a} and
the observation of magnetic dipole-dipole interaction (MDDI) in the
BEC~\cite{Stuhler:2005a}. In the latter one, we have shown that
depending on the orientation of the magnetic moments of the
condensed chromium atoms with respect to the long axis of our
optical dipole-trap, the expansion dynamics of the BEC is modified.
The dynamic behavior of the condensate aspect ratio after release
from an anisotropic trap was studied experimentally and compared to
numerical calculations based on the description of the chromium BEC
by superfluid hydrodynamic theory including dipole-dipole
interaction~\cite{Giovanazzi:2006,ODell:2004,Eberlein:2005}. The
observed behavior showed an excellent qualitative agreement with the
theoretical prediction. In this paper, we discuss a method that
allows to determine the strength of the MDDI compared to the contact
interaction with an accuracy on the percent level.\\
Since the absolute strength of the MDDI is known, one can use a
measurement of the relative strength of the MDDI to determine the
s-wave interaction strength, which is proportional to the s-wave
scattering length $a$. Many different techniques have been used to
determine the s-wave scattering lengths of ultra-cold atoms but most
of them come along with large error bars -- often because the number
of atoms enters the measurement. Examples are the $\rm ^{23}Na$
scattering length in $\ket{F=1,m_F=-1}$ of $a_{\rm Na}=92\pm25$
(i.e. $27\%$ error) determined from thermalization
measurements~\cite{Davis:1994}, and $a_{\rm Na}=65\pm30$ (i.e.
$46\%$ error) from the measurement of the mean field energy of a
BEC~\cite{Mewes:96}. The scattering length of metastable He$^*$
\unit[$a_{\rm He^*}= 16\pm 8$]{nm} was determined from the mean
field energy by analyzing the size of the BEC~\cite{Pereira:2001}
where the error of $50\%$ stems from an uncertainty of the number of
atoms in the condensate. Our first determination of the chromium
scattering length~\cite{Schmidt:2003b} was based on
cross-dimensional thermalization measurements~\cite{Monroe:1993} and
resulted in $a_{\rm Cr}=$170$\pm$39a$_0$. The error was mostly due
to an uncertainty in the density and atom-number determination. We
will show in this paper, that a measurement of the relative strength
of the magnetic dipole-dipole interaction in a BEC can be used to
obtain precise values for $a$ without such a strong dependence on
the determination of the number of atoms.\\
The interaction energy of two magnetic dipoles separated by the
distance $\vec{r}$ is given by
\[
\label{eq:diptheo:Udd} U_{dd}(\vec{r})=\frac{\mu_0 \mu_m^2}{4 \pi
r^3} \left( 1- \frac{3 (\vec{e}_\mu \vec{r} )^2}{r^2} \right)
\]
where the strength of the dipole-dipole interaction is measured by
the pre-factor of $U_{dd}$ and the orientation of the dipoles
$\vec{e}_\mu$ is parallel to an external magnetic field $\vec{B}$.
This strength can be compared to the coupling constant $g$ of the
s-wave interaction
\[
\label{eq:bectheo:pseudopot} U_{ sw}(\vec{r})=g
\delta(\vec{r})=\frac{4 \pi \hbar^2a}{m} \delta(\vec{r})
\]
and is measured by the dimensionless dipole-dipole strength
parameter
\[
\label{eq:diptheo:epsdd} \varepsilon_{
dd}=\frac{\mu_0\mu_m^2m}{12\pi\hbar^2a}.
\]
It is chosen such that a homogeneous condensate is unstable if
$\varepsilon_{ dd}>1$ in a static magnetic
field~\cite{Giovanazzi:2002a}.\\
In contrast to the s-wave interaction which can be understood as a
local, contact-like interaction (\ref{eq:bectheo:pseudopot}), the
dipole-dipole interaction is long-range and anisotropic. In a
condensate with density distribution
$n(\vec{r})=|\phi(\vec{r}\;)|^2$, it gives rise to the mean-field
potential~\cite{Goral:2000,ODell:2004}
\[
\label{eq:diptheo:Phidd} \Phi_{ dd}(\vec{r})=\int
U_{dd}(\vec{r}-\vec{r}\;')|\phi(\vec{r}\;')|^2 d^3r\;'.
\]
The integral in (\ref{eq:diptheo:Phidd}) reflects the non-local
character of the interaction. If this interaction in addition to the
contact interaction is taken into account, the well known
Gross-Pitaevskii equation gets the form
\begin{widetext}
\[
\label{eq:diptheo:GPEdip} i\hbar\frac{\partial}{\partial t}
\phi(\vec{r},t) = \left( -\frac{\hbar^2}{2m} \nabla^2 +
U_{ext}(\vec{r}) + g |\phi(\vec{r},t)|^2 + \int
U_{dd}(\vec{r}-\vec{r}\;') |\phi(\vec{r}\;',t)|^2 d^3{r}\;' \right)
\phi(\vec{r},t).
\]
\end{widetext}
O'Dell et al. have shown in~\cite{ODell:2004}, that even under the
influence of the dipole-dipole mean field potential
$\Phi_{dd}(\vec{r})$, the density distribution has the shape of an
inverted parabola in the Thomas-Fermi limit. Like in the case of
pure contact interaction, a wave function of the form
\[
|\phi(\vec{r})|^2=n_{c,0}\left(1-\frac{x^2}{R_x^2}-\frac{y^2}{R_y^2}-\frac{z^2}{R_z^2}\right)
\]
is a self consistent solution of the superfluid hydrodynamic
equations~\cite{Dalfovo:1999} derived from the Gross-Pitaevskii
equation~ (\ref{eq:diptheo:GPEdip}), even in presence of
dipole-dipole interaction~\cite{ODell:2004, Eberlein:2005}. $R_x$,
$R_y$, and $R_z$ are the Thomas-Fermi radii of the condensate. The
anisotropy of the dipole-dipole interaction manifests itself in a
modification of the aspect ratio of the trapped
condensate~\cite{Giovanazzi:2002a,Eberlein:2005}. This anisotropy
also reveals during the expansion of a dipolar
condensate~\cite{Giovanazzi:2003a,Stuhler:2005a}.\\
In the following we will determine the dipole-dipole strength
parameter $\varepsilon_{dd}$ by analyzing the dynamic behavior of
the Thomas-Fermi radii $R(t)$ of expanding dipolar condensates. The
experimental apparatus and techniques that are used are described in
detail in~\cite{Griesmaier:2006a,Giovanazzi:2006}. By applying a
small homogeneous external field ($\sim$\unit[11.5]{G}), oriented
either along the $y-$ or $z-$axis, shortly ($\sim$\unit[7]{ms})
before releasing the condensate from the trap, we obtained two sets
of measured radii of a ballistically expanding condensate with
different alignment of the atomic magnetic moments. The trap from
which the condensate was released was a crossed optical dipole trap
that was elongated in $z$-direction with trap parameters of
$\omega_x =2\pi\;$\unit[942]{Hz}, $\omega_y=2\pi\;$\unit[712]{Hz},
and $\omega_z=2\pi\;$\unit[128]{Hz}. In the asymptotic limit of long
times of flight, which is governed by a collisionless and potential
free (except for gravity) ballistic flight, the radii of the cloud
can be parameterized as
\[
\label{eq:dipexp:rt} R_{i}(t)=R_{i}^*+v_{i}^{*} t,
\]
where the index $i=[x,y,z]$ indicates the direction of expansion
that is considered. It is worthwhile to mention that the initial
values $R_{i}(0)=R_{i}^*$ are not the Thomas-Fermi radii $R_{i}$.
Note that the radii $R_i(t)$ as well as the asymptotic velocities
$v_{i}^{*}$ of the expansion for long times ($t \gg 1/\omega$)
depend on the direction in which the atoms are polarized. As shown
in~\cite{Giovanazzi:2006}, they are thus proportional to $(N
a)^{1/5}$. In particular,
\[
\label{eq:dipexp:vassymscale} v_{i}^{*}= C (N a)^{1/5}
\]
with a constant of proportionality $C$ that only depends on the
known or measured quantities that determine the chemical potential,
i.e. the trap parameters $\omega_x,\omega_y,\omega_z$ the atomic
mass $m$ and a small contribution of $\varepsilon_{dd}$. Using the
hydrodynamic theory of an expanding dipolar
condensate~\cite{Giovanazzi:2006}, the asymptotic velocity for a
certain number of atoms and scattering length can be easily
calculated numerically.
\begin{table}[t]
\centering
\begin{tabular}{ccc}
\hline \hline
polarization & $v_{y}^{*} [10^{-3} m/s]$ & $C\; [m^{4/5}] $ \\
\hline
no dipoles  & $8.528$  & $0.0488 $  \\
$y-$polarization    & $9.085$  & $0.0519  $  \\
$z-$polarization    & $8.283$  & $0.0474 $  \\
\hline
\end{tabular}
\caption{\label{tab:dipexp:propconst}Asymptotic velocity in
$y-$direction and corresponding proportionality constant $C$
calculated numerically for the case of vanishing dipole-dipole
interaction $\varepsilon_{dd}=0$, and for $\varepsilon_{dd}=0.148$
and polarization along $\hat{y}$ and $\hat{z}$. Velocities
calculated for $30000$ atoms, and $a=103 a_0$.}
\end{table}
Table~\ref{tab:dipexp:propconst} shows the expected asymptotic
velocities $v_{y}^{*}= C\cdot (30000 \cdot103 a_0)^{1/5}$ in
$y-$direction and the corresponding values for $C$, calculated for
pure contact interaction ($\varepsilon_{dd}=0$), $y-$polarization
and $z-$polarization. The numbers are calculated for the measured
trap parameters, $30000$ atoms, and a scattering length of $a=103
a_0$~\cite{Werner:2005}. The scattering length of $103 a_0$
corresponds to a dipole-dipole strength parameter of
$\varepsilon_{dd}=0.148$.\\
\begin{figure}
\centering
\begin{minipage}[t]{1\linewidth}
\centering
\includegraphics[width=0.49\columnwidth]{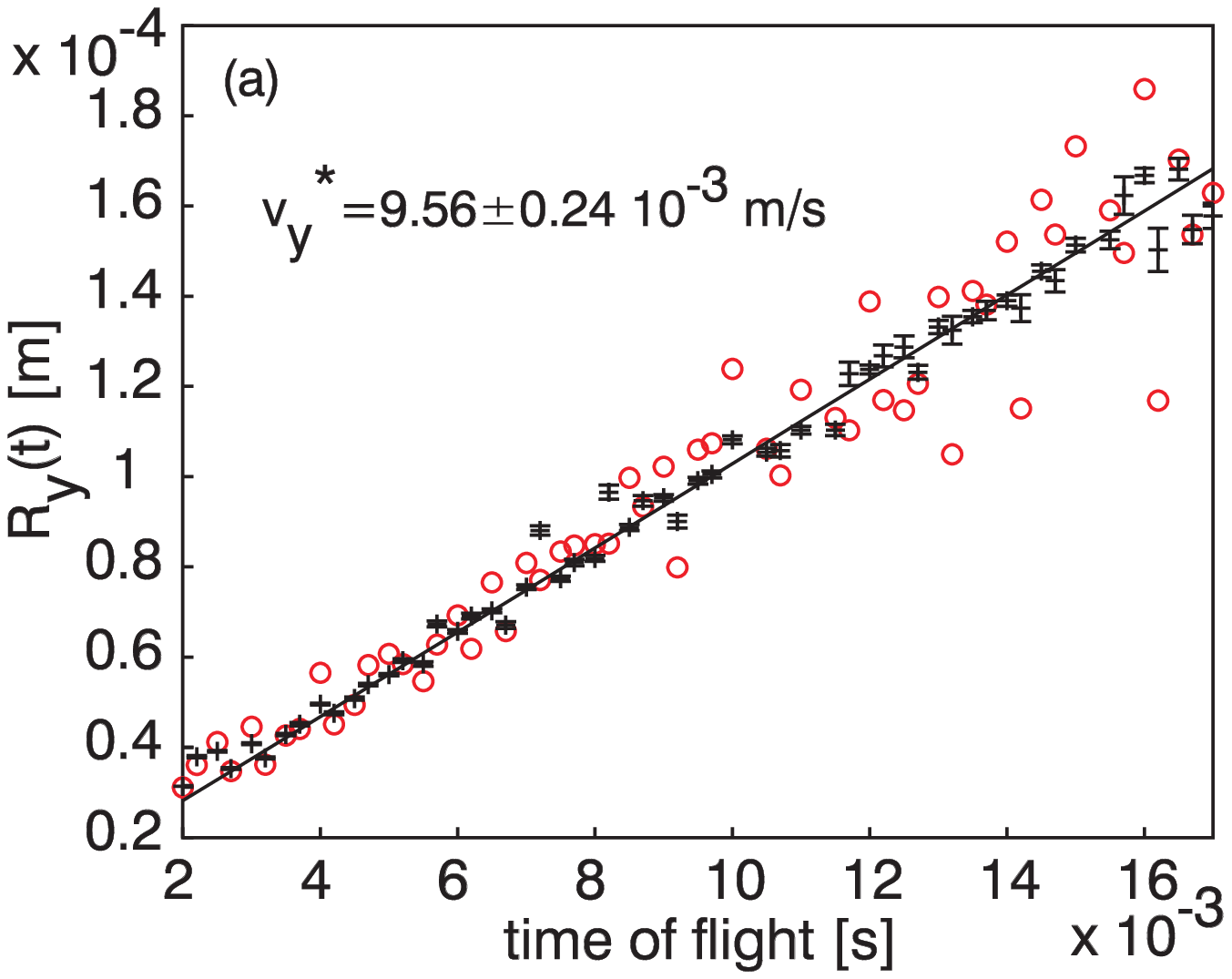}
\centering
\includegraphics[width=0.49\columnwidth]{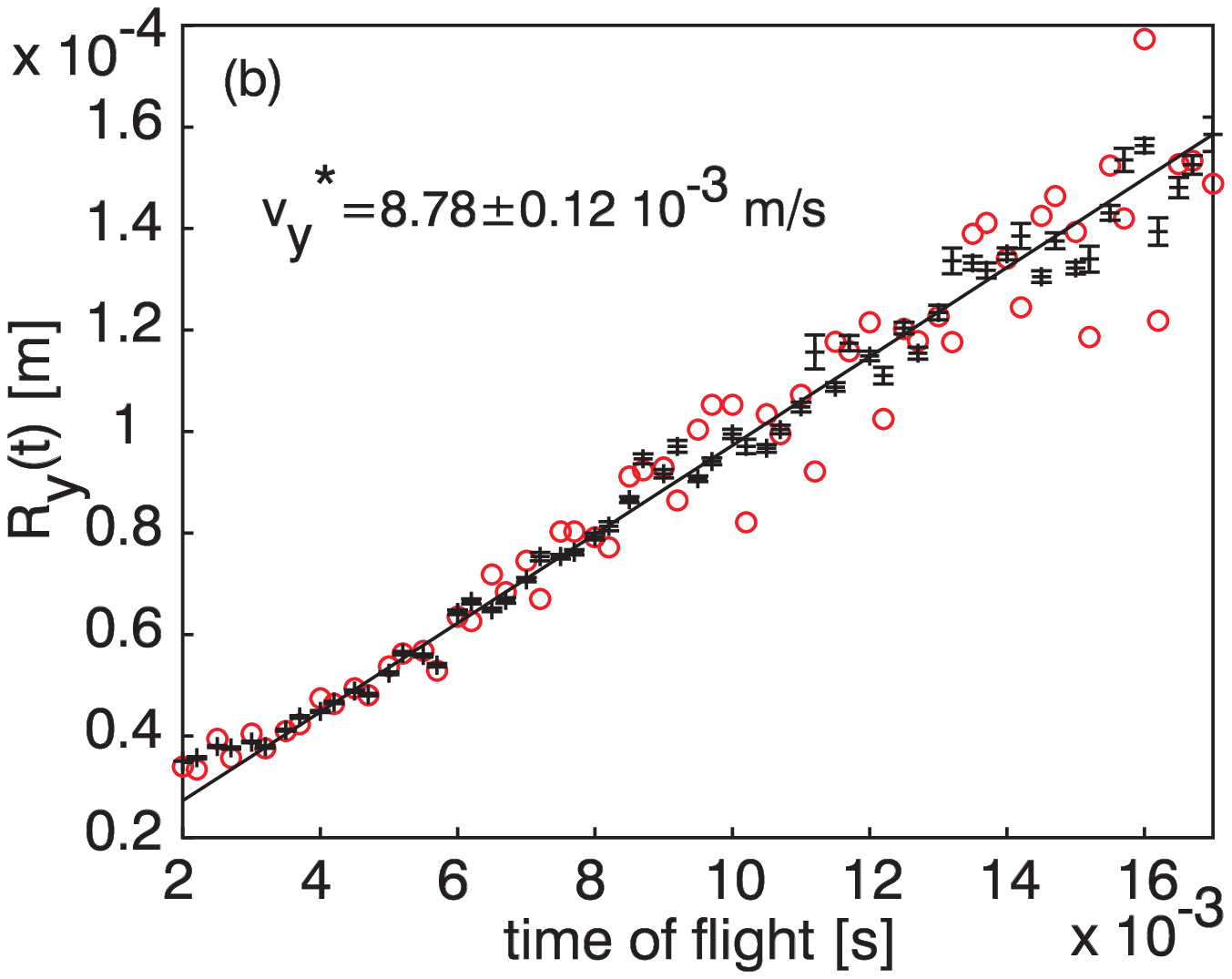}
\end{minipage}
\caption{\label{fig:dipexp:ddRy} Measured dependence of the
condensate radius $R_{y}(t)$ on the time of flight. (a) dipoles
aligned along the $y-$axis $(\vec{B}\|\hat{y})$, average number of
atoms was $29000$ (b) dipoles aligned along the $z-$axis
$(\vec{B}\|\hat{z})$, average number of atoms was $31000$. Open
circles: measured radii, crosses: measured radii re-scaled using
(\ref{eq:dipex:rescaler}); solid black line: linear fit to the
re-scaled radii.}
\end{figure}
To determine the asymptotic velocity, we use the condensate radii
$R(t)$ measured in a time-of-flight series. Since the number of
camera pixels that are covered by the condensate is much larger in
the direction of fast expansion ($y-$direction in our setup), one
can expect the most accurate results when considering $R_y(t)$.
Figure~\ref{fig:dipexp:ddRy} shows the dependence of the condensate
radius $R_y(t)$ with polarization along $\hat{y}$ (left figure) and
$\hat{z}$ (right figure) on the time of flight for $67$ different
expansion times. Because the number of atoms is fluctuating during
such a series of experiments, the radii fluctuate due to their
dependence on $N$. To get rid of these fluctuations, we divide each
measured radius $R_y$ by the fifth root of the number of condensed
atoms $N$ in the corresponding experiment and multiply them with the
mean value of the fifth root of the atom-numbers $<N^{1/5}>$ in all
experiments.
\[
\label{eq:dipex:rescaler}
\overline{R}_y=\frac{R_y}{N^{1/5}}<N^{1/5}>.
\]
In this way we get a series of time dependent radii which are now
independent of the atom-number. Open circles in
Fig.~\ref{fig:dipexp:ddRy} represent the measured $R_y(t)$, crosses
with error-bars mark the re-scaled $\overline{R}(t)$ which show much
less fluctuations. A linear fit to the re-scaled data for times
larger than \unit[3]{ms} to focus only on the asymptotic behavior,
yields $v_{y}^{*}=9.56\pm0.24$\unit{m/s} for $y-$polarization. For
$z-$polarization, we get $v_{y}^{*}=8.78\pm0.12$\unit{m/s}.\\
By using the above re-scaling, the errors $\Delta
v_{y}^{*}=\pm0.24$\unit{m/s} and $\Delta
v_{y}^{*}=\pm0.12$\unit{m/s} in the fitted slope $v_{y}^{*}$ for
$y-$ and $z-$polarization, respectively, do not contain fluctuations
of the atom-number anymore.\\
If we consider only the case of $y-$polarization, invert equation
(\ref{eq:dipexp:vassymscale}) and insert the fitted velocity
$v_{y}^{*}$ and the constant $C_y$ from
table~\ref{tab:dipexp:propconst}, we get
\begin{equation}
a=\frac{1}{<N>}\left(\frac{v_{y}^{*}}{C}\right)^5=138 a_0
\end{equation}
for the scattering length. The error on this measurement consists of
two contributions: first the fitted asymptotic velocity comes with
an error $\Delta v^{*}$ due to such kind of noise on the data that
is not correlated with the number of atoms. Since $v^{*}$ appears in
the fifth power in $a$, $\Delta v^{*}$ appears with a factor of $5$
in $\Delta a$. Second, the mean value of the number of atoms has an
uncertainty, mainly due to an uncertainty in the detuning of the
probe beam. Since a detuning from resonance can only lead to an
underestimation of the number of atoms, the error in the scattering
length $a$ caused by this uncertainty is only towards smaller values
of $a$. We estimate a maximum detuning of $\Delta
\delta_{probe}=\pm0.25\Gamma$ which leads to an estimated error in
the number of atoms of $\Delta N/N= - 0.25\%$. The relative error in
$a$ is then
\begin{equation}
\frac{\Delta a}{a}=\frac{\Delta <N>}{<N>}+5\frac{\Delta
v^{*}}{v^{*}}=-0.25 \pm 0.075.
\end{equation}
Hence the scattering length of $\rm ^{\rm 52}Cr$ determined with
this method is
\begin{equation}
a_{\rm Cr}=\Bigl( 138 {+10\atop-45} \Bigr) a_0.
\end{equation}
For $z$-polarization, we get a consistent value of  $a_{\rm
Cr}=\Bigl( 133 {+10\atop-43} \Bigr) a_0.$ Due to the relatively
large systematic error in the number of atoms, this way to determine
the scattering length yields only quite inaccurate values, typical
for condensate expansion experiments. In the following, we will use
the full set of data from both polarisations to determine the
scattering length with much higher
accuracy and independent of the number of condensed atoms.\\
We use the two rescaled asymptotic velocities \[
\label{eq:dipexp:vtilde}
\widetilde{v}_y^{*}=\frac{{v}_y^{*}}{<N^{1/5}>}
\]
${\widetilde{v}_y^{*}(\vec{B}\|\hat{y})}$ and
${\widetilde{v}_y^{*}(\vec{B}\|\hat{z})}$ for polarization along
$\hat{y}$ and $\hat{z}$, respectively to determine
$\varepsilon_{dd}$ by analyzing their ratio. To first order in
$\varepsilon_{dd}$ (in the expected range of $\varepsilon_{dd}$,
higher orders are negligible), the ratio has the form
\[
\label{eq:dipexp:vassymratio}
\frac{\widetilde{v}_y^{*}\;(\vec{B}\|\hat{y})}{\widetilde{v}_y^{*}\;(\vec{B}\|\hat{z})}=
1 + D\varepsilon_{dd}.
\]
It depends only on the asymmetry introduced by the dipole-dipole
interaction because the contribution of the s-wave scattering to the
total energy is independent of the polarization. $D$ is again a
numerical constant. If we use the measured asymptotic velocities, we
obtain
\begin{equation}
\varepsilon_{dd}=0.159\pm0.034
\end{equation}
in very good agreement with the value of $\varepsilon_{dd}=0.148$
that one would expect for $a_{Cr}=103a_0$.\\
%
%
\begin{table*}[t]
\begin{tabular}{cccc}
\hline \hline
method & year & scattering length $a_{Cr}$ $[a_0]$ & relative error\\
\hline
cross-dimensional thermalization~\cite{Schmidt:2003b} & 2003 & $170\pm39$ & $23\%$\\
Feshbach resonances~\cite{Werner:2005}   & 2005  & $103\pm13$ & $13\%$ \\
condensate expansion in one direction (this paper)& 2006 & $ 138
{+10\atop-45} a_0$ {and} $
133 {+10\atop-43}  a_0$& ${+8\atop-33}\%$\\
dipolar expansion (this paper) & 2006  & $96\pm20 $ & $13\%$ \\
\hline
\end{tabular}
\caption{\label{tab:dipexp:comparea} Comparison of experimental
values of the $^{ 52}Cr$ s-wave scattering length.}
\end{table*}
In turn, since the dipole-dipole interaction strength can be exactly
calculated from (\ref{eq:diptheo:epsdd}), this result can be used to
determine the scattering length:
\begin{equation*}
a=\frac{\mu_{ 0}\mu_{ m}^2m}{12\pi\hbar^2\varepsilon_{ dd}^{ meas}}=
96\pm20\;a_{ 0}.
\end{equation*}
This result is in excellent agreement with the value of
$103\pm13a_0$ that has been obtained by comparing the measured
positions of Feshbach resonances in chromium collisions with
multichannel calculations~\cite{Werner:2005}. Furthermore, the
relative error of $\pm20\%$ makes this way to determine the
scattering length a comparably accurate method.\\
Table~\ref{tab:dipexp:comparea} shows a comparison of the
values of the s-wave scattering length of chromium measured with different methods.\\
%
In conclusion, we have measured the relative strength of the
magnetic dipole-dipole interaction in a chromium BEC. The relative
strength parameter was measured $\varepsilon_{dd}=0.159\pm0.034$ by
analyzing the expansion of a chromium condensate with different
polarization after release from an anisotropic trapping potential.
This result was used to determine the s-wave scattering length of
$^{52}$Cr $a= 96\pm20\;a_0$ in excellent agreement with the results
of theoretical analysis of measured Feshbach resonances
($a_{^{52}Cr}=103\pm13a_0$)~\cite{Werner:2005}. In contrast to many
other methods that are commonly used to determine the s-wave
scattering length, this method does not depend on the accuracy of
the atom-number determination. Furthermore, unlike other methods
that deliver results with the same or even better accuracy like
Feshbach resonance measurements or photoassociation
spectroscopy~\cite{Abraham:1995,Moal:2006,Weiner:1999}, it does not
require knowledge of any details of the molecular potentials. We
expect it to be well suited to determine the scattering length close
to a Feshbach resonance with high accuracy, especially for small
scattering lengths where the dipole-dipole interaction becomes as
important as the contact interaction. The excellent agreement
between experimental results and theory constitutes a confirmation
of the theoretical approach that is used to describe the dipolar
BEC.
\begin{acknowledgments}
We are pleased to acknowledge fruitful discussions with Luis Santos
and Paolo Pedri. The research presented in this paper was supported
by the SFB/TR21 and SPP1116 of the German Science Foundation (DFG)
and by the Landesstiftung Baden-W{\"u}rttemberg.
\end{acknowledgments}
\bibliography{epsdd-minimal}

\begin{thebibliography}{31}
\expandafter\ifx\csname natexlab\endcsname\relax\def\natexlab#1{#1}\fi
\expandafter\ifx\csname bibnamefont\endcsname\relax
  \def\bibnamefont#1{#1}\fi
\expandafter\ifx\csname bibfnamefont\endcsname\relax
  \def\bibfnamefont#1{#1}\fi
\expandafter\ifx\csname citenamefont\endcsname\relax
  \def\citenamefont#1{#1}\fi
\expandafter\ifx\csname url\endcsname\relax
  \def\url#1{\texttt{#1}}\fi
\expandafter\ifx\csname urlprefix\endcsname\relax\def\urlprefix{URL }\fi
\providecommand{\bibinfo}[2]{#2}
\providecommand{\eprint}[2][]{\url{#2}}

\bibitem[{\citenamefont{Goral et~al.}(2000)\citenamefont{Goral, Rzazewski, and
  Pfau}}]{Goral:2000}
\bibinfo{author}{\bibfnamefont{K.}~\bibnamefont{Goral}},
  \bibinfo{author}{\bibfnamefont{K.}~\bibnamefont{Rzazewski}},
  \bibnamefont{and} \bibinfo{author}{\bibfnamefont{T.}~\bibnamefont{Pfau}},
  \bibinfo{journal}{Physical Review A} \textbf{\bibinfo{volume}{61}},
  \bibinfo{pages}{051601(R)} (\bibinfo{year}{2000}).

\bibitem[{\citenamefont{Yi and You}(2000)}]{Yi:2000}
\bibinfo{author}{\bibfnamefont{S.}~\bibnamefont{Yi}} \bibnamefont{and}
  \bibinfo{author}{\bibfnamefont{L.}~\bibnamefont{You}},
  \bibinfo{journal}{Physical Review A} \textbf{\bibinfo{volume}{61}},
  \bibinfo{pages}{041604(R)} (\bibinfo{year}{2000}).

\bibitem[{\citenamefont{Giovanazzi et~al.}(2003)\citenamefont{Giovanazzi,
  G\"{o}rlitz, and Pfau}}]{Giovanazzi:2003a}
\bibinfo{author}{\bibfnamefont{S.}~\bibnamefont{Giovanazzi}},
  \bibinfo{author}{\bibfnamefont{A.}~\bibnamefont{G\"{o}rlitz}},
  \bibnamefont{and} \bibinfo{author}{\bibfnamefont{T.}~\bibnamefont{Pfau}},
  \bibinfo{journal}{J.~Opt.~B: Quantum Semiclass.~Opt.}
  \textbf{\bibinfo{volume}{5}}, \bibinfo{pages}{S208} (\bibinfo{year}{2003}).

\bibitem[{\citenamefont{Yi and You}(2003)}]{Yi:2003}
\bibinfo{author}{\bibfnamefont{S.}~\bibnamefont{Yi}} \bibnamefont{and}
  \bibinfo{author}{\bibfnamefont{L.}~\bibnamefont{You}},
  \bibinfo{journal}{Physical Review A} \textbf{\bibinfo{volume}{67}},
  \bibinfo{pages}{045601} (\bibinfo{year}{2003}).

\bibitem[{\citenamefont{Giovanazzi et~al.}(2006)\citenamefont{Giovanazzi,
  Pedri, Santos, Griesmaier, Fattori, Koch, Stuhler, and
  Pfau}}]{Giovanazzi:2006}
\bibinfo{author}{\bibfnamefont{S.}~\bibnamefont{Giovanazzi}},
  \bibinfo{author}{\bibfnamefont{P.}~\bibnamefont{Pedri}},
  \bibinfo{author}{\bibfnamefont{L.}~\bibnamefont{Santos}},
  \bibinfo{author}{\bibfnamefont{A.}~\bibnamefont{Griesmaier}},
  \bibinfo{author}{\bibfnamefont{M.}~\bibnamefont{Fattori}},
  \bibinfo{author}{\bibfnamefont{T.}~\bibnamefont{Koch}},
  \bibinfo{author}{\bibfnamefont{J.}~\bibnamefont{Stuhler}}, \bibnamefont{and}
  \bibinfo{author}{\bibfnamefont{T.}~\bibnamefont{Pfau}},
  \bibinfo{journal}{Physical Review A} \textbf{\bibinfo{volume}{74}},
  \bibinfo{pages}{013621} (\bibinfo{year}{2006}).

\bibitem[{\citenamefont{{Santos} et~al.}(2003)\citenamefont{{Santos},
  {Shlyapnikov}, and {Lewenstein}}}]{Santos:2003}
\bibinfo{author}{\bibfnamefont{L.}~\bibnamefont{{Santos}}},
  \bibinfo{author}{\bibfnamefont{G.~V.} \bibnamefont{{Shlyapnikov}}},
  \bibnamefont{and}
  \bibinfo{author}{\bibfnamefont{M.}~\bibnamefont{{Lewenstein}}},
  \bibinfo{journal}{Physical Review Letters} \textbf{\bibinfo{volume}{90}},
  \bibinfo{pages}{250403} (\bibinfo{year}{2003}).

\bibitem[{\citenamefont{Goral and Santos}(2002)}]{Goral:2002}
\bibinfo{author}{\bibfnamefont{K.}~\bibnamefont{Goral}} \bibnamefont{and}
  \bibinfo{author}{\bibfnamefont{L.}~\bibnamefont{Santos}},
  \bibinfo{journal}{Physical Review A} \textbf{\bibinfo{volume}{66}},
  \bibinfo{pages}{023613} (\bibinfo{year}{2002}).

\bibitem[{\citenamefont{O'Dell et~al.}(2004)\citenamefont{O'Dell, Giovanazzi,
  and Eberlein}}]{ODell:2004}
\bibinfo{author}{\bibfnamefont{D.~H.~J.} \bibnamefont{O'Dell}},
  \bibinfo{author}{\bibfnamefont{S.}~\bibnamefont{Giovanazzi}},
  \bibnamefont{and} \bibinfo{author}{\bibfnamefont{C.}~\bibnamefont{Eberlein}},
  \bibinfo{journal}{Physical Review Letters} \textbf{\bibinfo{volume}{92}},
  \bibinfo{pages}{250401} (\bibinfo{year}{2004}).

\bibitem[{\citenamefont{Pedri and Santos}(2005)}]{Pedri:2005}
\bibinfo{author}{\bibfnamefont{P.}~\bibnamefont{Pedri}} \bibnamefont{and}
  \bibinfo{author}{\bibfnamefont{L.}~\bibnamefont{Santos}},
  \bibinfo{journal}{Physical Review Letters} \textbf{\bibinfo{volume}{95}},
  \bibinfo{pages}{200404} (\bibinfo{year}{2005}).

\bibitem[{\citenamefont{Santos et~al.}(2000)\citenamefont{Santos, Shlyapnikov,
  Zoller, and Lewenstein}}]{Santos:2000}
\bibinfo{author}{\bibfnamefont{L.}~\bibnamefont{Santos}},
  \bibinfo{author}{\bibfnamefont{G.~V.} \bibnamefont{Shlyapnikov}},
  \bibinfo{author}{\bibfnamefont{P.}~\bibnamefont{Zoller}}, \bibnamefont{and}
  \bibinfo{author}{\bibfnamefont{M.}~\bibnamefont{Lewenstein}},
  \bibinfo{journal}{Physical Review Letters} \textbf{\bibinfo{volume}{85}},
  \bibinfo{pages}{1791} (\bibinfo{year}{2000}).

\bibitem[{\citenamefont{G{\'o}ral et~al.}(2002)\citenamefont{G{\'o}ral, Santos,
  and Lewenstein}}]{Goral:2002a}
\bibinfo{author}{\bibfnamefont{K.}~\bibnamefont{G{\'o}ral}},
  \bibinfo{author}{\bibfnamefont{L.}~\bibnamefont{Santos}}, \bibnamefont{and}
  \bibinfo{author}{\bibfnamefont{M.}~\bibnamefont{Lewenstein}},
  \bibinfo{journal}{Physical~Review~Letters} \textbf{\bibinfo{volume}{88}},
  \bibinfo{pages}{170406} (\bibinfo{year}{2002}).

\bibitem[{\citenamefont{Cooper et~al.}(2005)\citenamefont{Cooper, Rezayi, and
  Simon}}]{Cooper:2005}
\bibinfo{author}{\bibfnamefont{N.~R.} \bibnamefont{Cooper}},
  \bibinfo{author}{\bibfnamefont{E.~H.} \bibnamefont{Rezayi}},
  \bibnamefont{and} \bibinfo{author}{\bibfnamefont{S.~H.} \bibnamefont{Simon}},
  \bibinfo{journal}{Physical Review Letters} \textbf{\bibinfo{volume}{95}},
  \bibinfo{pages}{200402} (\bibinfo{year}{2005}).

\bibitem[{\citenamefont{Rezayi et~al.}(2005)\citenamefont{Rezayi, Read, and
  Cooper}}]{Rezayi:2005}
\bibinfo{author}{\bibfnamefont{E.~H.} \bibnamefont{Rezayi}},
  \bibinfo{author}{\bibfnamefont{N.}~\bibnamefont{Read}}, \bibnamefont{and}
  \bibinfo{author}{\bibfnamefont{N.~R.} \bibnamefont{Cooper}},
  \bibinfo{journal}{Physical Review Letters} \textbf{\bibinfo{volume}{95}},
  \bibinfo{pages}{160404} (\bibinfo{year}{2005}).

\bibitem[{\citenamefont{Santos and Pfau}(2006)}]{Santos:2005a}
\bibinfo{author}{\bibfnamefont{L.}~\bibnamefont{Santos}} \bibnamefont{and}
  \bibinfo{author}{\bibfnamefont{T.}~\bibnamefont{Pfau}},
  \bibinfo{journal}{Physical Review Letters} \textbf{\bibinfo{volume}{96}},
  \bibinfo{pages}{190404} (\bibinfo{year}{2006}).

\bibitem[{\citenamefont{Diener and Ho}(2006)}]{Diener:2005}
\bibinfo{author}{\bibfnamefont{R.~B.} \bibnamefont{Diener}} \bibnamefont{and}
  \bibinfo{author}{\bibfnamefont{T.-L.} \bibnamefont{Ho}},
  \bibinfo{journal}{Physical Review Letters} \textbf{\bibinfo{volume}{96}},
  \bibinfo{pages}{190405} (\bibinfo{year}{2006}).

\bibitem[{\citenamefont{Kawaguchi et~al.}(2006)\citenamefont{Kawaguchi, Saito,
  and Ueda}}]{Kawaguchi:2005}
\bibinfo{author}{\bibfnamefont{Y.}~\bibnamefont{Kawaguchi}},
  \bibinfo{author}{\bibfnamefont{H.}~\bibnamefont{Saito}}, \bibnamefont{and}
  \bibinfo{author}{\bibfnamefont{M.}~\bibnamefont{Ueda}},
  \bibinfo{journal}{Physical Review Letters} \textbf{\bibinfo{volume}{96}},
  \bibinfo{pages}{080405} (\bibinfo{year}{2006}).

\bibitem[{\citenamefont{Griesmaier et~al.}(2005)\citenamefont{Griesmaier,
  Werner, Hensler, Stuhler, and Pfau}}]{Griesmaier:2005a}
\bibinfo{author}{\bibfnamefont{A.}~\bibnamefont{Griesmaier}},
  \bibinfo{author}{\bibfnamefont{J.}~\bibnamefont{Werner}},
  \bibinfo{author}{\bibfnamefont{S.}~\bibnamefont{Hensler}},
  \bibinfo{author}{\bibfnamefont{J.}~\bibnamefont{Stuhler}}, \bibnamefont{and}
  \bibinfo{author}{\bibfnamefont{T.}~\bibnamefont{Pfau}},
  \bibinfo{journal}{Physical~Review~Letters} \textbf{\bibinfo{volume}{94}},
  \bibinfo{pages}{160401} (\bibinfo{year}{2005}).

\bibitem[{\citenamefont{Stuhler et~al.}(2005)\citenamefont{Stuhler, Griesmaier,
  Koch, Fattori, Pfau, Giovanazzi, Pedri, and Santos}}]{Stuhler:2005a}
\bibinfo{author}{\bibfnamefont{J.}~\bibnamefont{Stuhler}},
  \bibinfo{author}{\bibfnamefont{A.}~\bibnamefont{Griesmaier}},
  \bibinfo{author}{\bibfnamefont{T.}~\bibnamefont{Koch}},
  \bibinfo{author}{\bibfnamefont{M.}~\bibnamefont{Fattori}},
  \bibinfo{author}{\bibfnamefont{T.}~\bibnamefont{Pfau}},
  \bibinfo{author}{\bibfnamefont{S.}~\bibnamefont{Giovanazzi}},
  \bibinfo{author}{\bibfnamefont{P.}~\bibnamefont{Pedri}}, \bibnamefont{and}
  \bibinfo{author}{\bibfnamefont{L.}~\bibnamefont{Santos}},
  \bibinfo{journal}{Physical Review Letters} \textbf{\bibinfo{volume}{95}},
  \bibinfo{pages}{150406} (\bibinfo{year}{2005}).

\bibitem[{\citenamefont{Eberlein et~al.}(2005)\citenamefont{Eberlein,
  Giovanazzi, and O'Dell}}]{Eberlein:2005}
\bibinfo{author}{\bibfnamefont{C.}~\bibnamefont{Eberlein}},
  \bibinfo{author}{\bibfnamefont{S.}~\bibnamefont{Giovanazzi}},
  \bibnamefont{and} \bibinfo{author}{\bibfnamefont{D.~H.~J.}
  \bibnamefont{O'Dell}}, \bibinfo{journal}{Physical Review A}
  \textbf{\bibinfo{volume}{71}}, \bibinfo{pages}{033618}
  (\bibinfo{year}{2005}).

\bibitem[{\citenamefont{Davis et~al.}(1995)\citenamefont{Davis, Mewes, Joffe,
  Andrews, and Ketterle}}]{Davis:1994}
\bibinfo{author}{\bibfnamefont{K.~B.} \bibnamefont{Davis}},
  \bibinfo{author}{\bibfnamefont{M.-O.} \bibnamefont{Mewes}},
  \bibinfo{author}{\bibfnamefont{M.~A.} \bibnamefont{Joffe}},
  \bibinfo{author}{\bibfnamefont{M.~R.} \bibnamefont{Andrews}},
  \bibnamefont{and} \bibinfo{author}{\bibfnamefont{W.}~\bibnamefont{Ketterle}},
  \bibinfo{journal}{Physical Review Letters} \textbf{\bibinfo{volume}{74}},
  \bibinfo{pages}{5202} (\bibinfo{year}{1995}).

\bibitem[{\citenamefont{Mewes et~al.}(1996)\citenamefont{Mewes, Andrews, van
  Druten, Kurn, Durfee, and Ketterle}}]{Mewes:96}
\bibinfo{author}{\bibfnamefont{M.-O.} \bibnamefont{Mewes}},
  \bibinfo{author}{\bibfnamefont{M.~R.} \bibnamefont{Andrews}},
  \bibinfo{author}{\bibfnamefont{N.~J.} \bibnamefont{van Druten}},
  \bibinfo{author}{\bibfnamefont{D.~M.} \bibnamefont{Kurn}},
  \bibinfo{author}{\bibfnamefont{D.~S.} \bibnamefont{Durfee}},
  \bibnamefont{and} \bibinfo{author}{\bibfnamefont{W.}~\bibnamefont{Ketterle}},
  \bibinfo{journal}{Physical~Review~Letters} \textbf{\bibinfo{volume}{77}},
  \bibinfo{pages}{416} (\bibinfo{year}{1996}).

\bibitem[{\citenamefont{PereiraDosSantos
  et~al.}(2001)\citenamefont{PereiraDosSantos, Leonard, Wang, Barrelet,
  Perales, Rasel, Unnikrishnan, Leduc, and Cohen-Tannoudji}}]{Pereira:2001}
\bibinfo{author}{\bibfnamefont{F.}~\bibnamefont{PereiraDosSantos}},
  \bibinfo{author}{\bibfnamefont{J.}~\bibnamefont{Leonard}},
  \bibinfo{author}{\bibfnamefont{J.}~\bibnamefont{Wang}},
  \bibinfo{author}{\bibfnamefont{C.~J.} \bibnamefont{Barrelet}},
  \bibinfo{author}{\bibfnamefont{F.}~\bibnamefont{Perales}},
  \bibinfo{author}{\bibfnamefont{E.}~\bibnamefont{Rasel}},
  \bibinfo{author}{\bibfnamefont{C.~S.} \bibnamefont{Unnikrishnan}},
  \bibinfo{author}{\bibfnamefont{M.}~\bibnamefont{Leduc}}, \bibnamefont{and}
  \bibinfo{author}{\bibfnamefont{C.}~\bibnamefont{Cohen-Tannoudji}},
  \bibinfo{journal}{Physical Review Letters} \textbf{\bibinfo{volume}{86}},
  \bibinfo{pages}{3459} (\bibinfo{year}{2001}).

\bibitem[{\citenamefont{Schmidt et~al.}(2003)\citenamefont{Schmidt, Hensler,
  Werner, Griesmaier, G\"{o}rlitz, Pfau, and Simoni}}]{Schmidt:2003b}
\bibinfo{author}{\bibfnamefont{P.~O.} \bibnamefont{Schmidt}},
  \bibinfo{author}{\bibfnamefont{S.}~\bibnamefont{Hensler}},
  \bibinfo{author}{\bibfnamefont{J.}~\bibnamefont{Werner}},
  \bibinfo{author}{\bibfnamefont{A.}~\bibnamefont{Griesmaier}},
  \bibinfo{author}{\bibfnamefont{A.}~\bibnamefont{G\"{o}rlitz}},
  \bibinfo{author}{\bibfnamefont{T.}~\bibnamefont{Pfau}}, \bibnamefont{and}
  \bibinfo{author}{\bibfnamefont{A.}~\bibnamefont{Simoni}},
  \bibinfo{journal}{Physical~Review~Letters} \textbf{\bibinfo{volume}{91}},
  \bibinfo{pages}{193201} (\bibinfo{year}{2003}).

\bibitem[{\citenamefont{Monroe et~al.}(1993)\citenamefont{Monroe, Cornell,
  Sackett, Myatt, and Wieman}}]{Monroe:1993}
\bibinfo{author}{\bibfnamefont{C.~R.} \bibnamefont{Monroe}},
  \bibinfo{author}{\bibfnamefont{E.~A.} \bibnamefont{Cornell}},
  \bibinfo{author}{\bibfnamefont{C.~A.} \bibnamefont{Sackett}},
  \bibinfo{author}{\bibfnamefont{C.~J.} \bibnamefont{Myatt}}, \bibnamefont{and}
  \bibinfo{author}{\bibfnamefont{C.~E.} \bibnamefont{Wieman}},
  \bibinfo{journal}{Physical Review Letters} \textbf{\bibinfo{volume}{70}},
  \bibinfo{pages}{414} (\bibinfo{year}{1993}).

\bibitem[{\citenamefont{Giovanazzi et~al.}(2002)\citenamefont{Giovanazzi,
  G\"{o}rlitz, and Pfau}}]{Giovanazzi:2002a}
\bibinfo{author}{\bibfnamefont{S.}~\bibnamefont{Giovanazzi}},
  \bibinfo{author}{\bibfnamefont{A.}~\bibnamefont{G\"{o}rlitz}},
  \bibnamefont{and} \bibinfo{author}{\bibfnamefont{T.}~\bibnamefont{Pfau}},
  \bibinfo{journal}{Physical~Review~Letters} \textbf{\bibinfo{volume}{89}},
  \bibinfo{pages}{130401} (\bibinfo{year}{2002}).

\bibitem[{\citenamefont{{Dalfovo} et~al.}(1999)\citenamefont{{Dalfovo},
  {Giorgini}, {Pitaevskii}, and {Stringari}}}]{Dalfovo:1999}
\bibinfo{author}{\bibfnamefont{F.}~\bibnamefont{{Dalfovo}}},
  \bibinfo{author}{\bibfnamefont{S.}~\bibnamefont{{Giorgini}}},
  \bibinfo{author}{\bibfnamefont{L.~P.} \bibnamefont{{Pitaevskii}}},
  \bibnamefont{and}
  \bibinfo{author}{\bibfnamefont{S.}~\bibnamefont{{Stringari}}},
  \bibinfo{journal}{Reviews of Modern Physics} \textbf{\bibinfo{volume}{71}},
  \bibinfo{pages}{463} (\bibinfo{year}{1999}).

\bibitem[{\citenamefont{Griesmaier et~al.}(2006)\citenamefont{Griesmaier,
  Stuhler, and Pfau}}]{Griesmaier:2006a}
\bibinfo{author}{\bibfnamefont{A.}~\bibnamefont{Griesmaier}},
  \bibinfo{author}{\bibfnamefont{J.}~\bibnamefont{Stuhler}}, \bibnamefont{and}
  \bibinfo{author}{\bibfnamefont{T.}~\bibnamefont{Pfau}},
  \bibinfo{journal}{Appl. Phys. B} \textbf{\bibinfo{volume}{82}},
  \bibinfo{pages}{211} (\bibinfo{year}{2006}).

\bibitem[{\citenamefont{Werner et~al.}(2005)\citenamefont{Werner, Griesmaier,
  Hensler, Stuhler, Pfau, Simoni, and Tiesinga}}]{Werner:2005}
\bibinfo{author}{\bibfnamefont{J.}~\bibnamefont{Werner}},
  \bibinfo{author}{\bibfnamefont{A.}~\bibnamefont{Griesmaier}},
  \bibinfo{author}{\bibfnamefont{S.}~\bibnamefont{Hensler}},
  \bibinfo{author}{\bibfnamefont{J.}~\bibnamefont{Stuhler}},
  \bibinfo{author}{\bibfnamefont{T.}~\bibnamefont{Pfau}},
  \bibinfo{author}{\bibfnamefont{A.}~\bibnamefont{Simoni}}, \bibnamefont{and}
  \bibinfo{author}{\bibfnamefont{E.}~\bibnamefont{Tiesinga}},
  \bibinfo{journal}{Physical~Review~Letters} \textbf{\bibinfo{volume}{94}},
  \bibinfo{pages}{183201} (\bibinfo{year}{2005}).

\bibitem[{\citenamefont{Abraham et~al.}(1995)\citenamefont{Abraham,
  McAlexander, Sackett, and Hulet}}]{Abraham:1995}
\bibinfo{author}{\bibfnamefont{E.~R.~I.} \bibnamefont{Abraham}},
  \bibinfo{author}{\bibfnamefont{W.~I.} \bibnamefont{McAlexander}},
  \bibinfo{author}{\bibfnamefont{C.~A.} \bibnamefont{Sackett}},
  \bibnamefont{and} \bibinfo{author}{\bibfnamefont{R.~G.} \bibnamefont{Hulet}},
  \bibinfo{journal}{Phys. Rev. Lett.} \textbf{\bibinfo{volume}{74}},
  \bibinfo{pages}{1315} (\bibinfo{year}{1995}).

\bibitem[{\citenamefont{Moal et~al.}(2006)\citenamefont{Moal, Portier, Kim,
  Dugue, Rapol, Leduc, and Cohen-Tannoudji}}]{Moal:2006}
\bibinfo{author}{\bibfnamefont{S.}~\bibnamefont{Moal}},
  \bibinfo{author}{\bibfnamefont{M.}~\bibnamefont{Portier}},
  \bibinfo{author}{\bibfnamefont{J.}~\bibnamefont{Kim}},
  \bibinfo{author}{\bibfnamefont{J.}~\bibnamefont{Dugue}},
  \bibinfo{author}{\bibfnamefont{U.~D.} \bibnamefont{Rapol}},
  \bibinfo{author}{\bibfnamefont{M.}~\bibnamefont{Leduc}}, \bibnamefont{and}
  \bibinfo{author}{\bibfnamefont{C.}~\bibnamefont{Cohen-Tannoudji}},
  \bibinfo{journal}{Physical Review Letters} \textbf{\bibinfo{volume}{96}},
  \bibinfo{pages}{023203} (\bibinfo{year}{2006}).

\bibitem[{\citenamefont{Weiner et~al.}(1999)\citenamefont{Weiner, Bagnato,
  Zilio, and Julienne}}]{Weiner:1999}
\bibinfo{author}{\bibfnamefont{J.}~\bibnamefont{Weiner}},
  \bibinfo{author}{\bibfnamefont{V.~S.} \bibnamefont{Bagnato}},
  \bibinfo{author}{\bibfnamefont{S.}~\bibnamefont{Zilio}}, \bibnamefont{and}
  \bibinfo{author}{\bibfnamefont{P.~S.} \bibnamefont{Julienne}},
  \bibinfo{journal}{Rev. Mod. Phys.} \textbf{\bibinfo{volume}{71}},
  \bibinfo{pages}{1} (\bibinfo{year}{1999}), \bibinfo{note}{and references
  therein}.

\end{thebibliography}
\end{document}